\documentclass[
reprint,nofootinbib,amsmath,amssymb,aps,preprintnumbers,superscriptaddress
]{revtex4-1}

\usepackage{graphicx}
\usepackage{bm}
\usepackage{braket}
\usepackage{cases} 
\usepackage{here}
\usepackage{color}
\usepackage{upgreek}
\usepackage{ulem}

\usepackage{dcolumn}

\begin{document}

\preprint{KOBE-COSMO-20-14}

\title{Analytic formula for the dynamics around
inflation end and implications on primordial gravitational waves}

\author{Asuka Ito}
\email[]{asuka-ito@godzilla.kobe-u.ac.jp}

\affiliation{Department of Physics, Kobe University, Kobe 657-8501, Japan}

\author{Jiro Soda }
\email[]{jiro@phys.sci.kobe-u.ac.jp}

\affiliation{Department of Physics, Kobe University, Kobe 657-8501, Japan}

\author{Masahide Yamaguchi}
\email[]{gucci@phys.titech.ac.jp}

\affiliation{Department of Physics, Tokyo Institute of Technology, Tokyo 152-8551, Japan}


\begin{abstract}
We argue that primordial gravitational waves have a spectral break and
its information is quite useful for exploring the early universe.
Indeed, such a spectral break can be a fingerprint of the end of
inflation, and the amplitude and the frequency at the break can tell us
the energy scale of inflation and the reheating temperature
simultaneously. In order to investigate the spectral break, we give an
analytic formula for evolution of the Hubble parameter around the end of
inflation where the slow roll approximation breaks down.  We also
evaluate the spectrum of primordial gravitational waves around the
break point semi-analytically using the analytic formula for the
inflation dynamics.
\end{abstract}

\maketitle

%
%
%
%
%
%
%
%
\section{Introduction}
Inflation, the accelerated expansion in the early universe, generates
primordial density and tensor perturbations as well as solving the
flatness and the horizon 
problems~\cite{Starobinsky:1980te,Sato:1980yn,Guth:1980zm}. 
The experiments of
cosmic microwave background (CMB) anisotropies have directly observed
primordial density perturbations and their amplitudes were shown to be
$\sim 10^{-5}$ on large scales~\cite{Smoot:1992td,Akrami:2018odb}, which yield important
information on the dynamics of inflation, that is, $\delta \rho/\rho
\sim H^2/\dot{\phi}$~\cite{Mukhanov:1981xt,Hawking:1982cz,Starobinsky:1982ee,Guth:1982ec}.
Unfortunately,
this quantity depends on the velocity of an inflaton as well as on the
Hubble parameter during inflation, which determines the energy scale of
inflation. 
Thus, the detection of primordial tensor perturbations,
whose amplitudes are solely given by the Hubble 
parameter~\cite{Grishchuk:1974ny,Starobinsky:1979ty}, is pressing. 
The primordial tensor
perturbations (gravitational waves) can be probed in two ways. One way
is to observe them through the B-mode polarizations of the CMB
indirectly~\cite{Akrami:2018odb,Ade:2015tva,Abazajian:2016yjj,Hazumi:2019lys}. 
The other way is to directly detect them through, for example,
interferometer experiments~\cite{LIGO,Virgo}. 
In fact, recent observations of the
gravitational waves sourced by binary blackholes through the LIGO/VIRGO
collaborations open gravitational wave astronomy~\cite{Abbott:2016blz}. 
There are several ongoing and planned experiments
to directly detect the primordial gravitational 
waves~\cite{Kawamura_2011,Somiya:2011np,2017arXiv170200786A}.

The primordial gravitational waves produced during inflation have
fruitful information on the dynamics of inflation.%
\footnote{Sourced gravitational waves also can be produced 
during inflaton in addition to the vacuum 
fluctuations~\cite{Senatore:2011sp,Fujita:2017jwq}.
We do not consider such possibilities because they are model-dependent.
We only consider single field inflation models to treat issues in a
model-independent way.}
Since they have
almost scale invariant spectrum on horizon exit in general, their
amplitudes give us the energy scale of inflation. 
While they are frozen and
their amplitudes are kept on superhorizon scales, they behave as damped
oscillations on subhorizon scales. Thus, the (would-be) observed
spectrum of primordial gravitational waves depend on cosmic history
much. That is, even if they were exactly scale invariant when they exited the
horizon, the (would-be) observed one can be scale-dependent. By use of
this kind of scale dependence, one can probe the change of the number of
relativistic degrees of freedom~\cite{Watanabe:2006qe,Kuroyanagi:2008ye}, 
the reheating temperature of the
universe~\cite{Nakayama:2008ip,Nakayama:2008wy,Kuroyanagi:2014qza}, for example.

An important feature in the spectrum of primordial gravitational waves is 
a break due to the end of inflation.
Indeed, we will show that the frequency and the amplitude at the break tells
us the energy scale of inflation and the reheating temperature simultaneously.
So far, rough analytic 
estimates~\cite{Maggiore:1999vm} 
or detailed numerical 
calculations~\cite{Kuroyanagi:2008ye} have been done
to clarify the spectral shape of the (would-be) observed primordial
gravitational waves around the break point. 
However, in order to properly incorporate the history of reheating 
in the analysis of the spectrum, 
it is desired to have a more accurate analytic formula.
For example, Ema {\it et
al.} recently pointed out that, even after inflation end, 
primordial gravitational waves might be produced as a result of 
inflaton annihilation into graviton 
pairs~\cite{Ema:2020ggo}.
Thus, in order to smoothly connect the
spectral shape at the frequency which exits the horizon just at the end
of inflation, we need the detailed information around this frequency. 
For this purpose, we will derive an analytic formula to
approximate the dynamics of inflation around its end, which also
yields a semi-analytic formula of the spectral of the primordial
gravitational waves around the frequency which exits the horizon at the
end of inflation.
It should be noted that, typically, the frequency of the spectral break is around $10^{7}$ Hz
(see Eq.\,(\ref{fcut1})).
In this frequency range, new gravitational wave detectors are being proposed and 
developed intensively~\cite{Domcke:2020yzq,Ito:2020wxi,Ito:2019wcb,Li:2009zzy,Li:2008qr}.
Therefore, our study is also important for the future gravitational wave experiments.

The paper is organized as follows. 
In Sec.\,\ref{sec2}, after giving the
basic expression of the amplitude of primordial gravitational waves
generated during inflation, we explain how to determine the Hubble
parameter and the reheating temperature from the information of
primordial gravitational waves on the break scale. 
In Sec.\,\ref{HuEvo}, an
analytic formula for evolution of the Hubble parameter around the end of 
inflation is derived. 
In Sec.\,\ref{PSpe}, 
we first discuss the spectral index of the spectrum at the end of inflation.
An implication to reconstruct the inflaton potential from the observation of the spectral index
with our formula is given.
We also semi-analytically give an example of the spectrum around the break point at present.
Final section is devoted to conclusion.
\section{The spectral break}\label{sec2}
Inflation predicts a nearly scale invariant spectrum of primordial gravitational 
waves~\cite{Grishchuk:1974ny,Starobinsky:1979ty}.
More explicitly, the dimensionless power spectrum is given by 
\begin{equation}
  P_{h}(k) = \frac{2}{\pi^{2}}\frac{H^{2}}{M_{\rm pl}^{2}}\Big|_{k=aH} \ ,  \label{spe}
\end{equation}
where  $H$ is the Hubble parameter, $M_{\rm pl}$ is the reduced Planck mass, and $a$ is the scale factor.
Here, we evaluated the power spectrum at the horizon crossing, $k=aH$.
The amplitude of the power spectrum is determined by $H$ and nearly scale invariant because $H$ is almost constant during inflation.
Therefore, if we observe the scale invariant spectrum of primordial gravitational waves,
we see the energy scale of inflation through the parameter $H$.
Indeed, upper bounds on the energy scale of inflation at the pivot scale is given by the observation of 
cosmic microwave background~\cite{Akrami:2018odb}.

A feature in the primordial gravitational wave spectrum produced during inflation is 
the break around the end of inflation. 
Above the break frequency, production of primordial gravitational waves should be exponentially suppressed though Ema {\it et
al.} recently pointed out that, even above such break frequency, 
the production of primordial gravitational waves might happen as a result of 
inflaton annihilation into graviton pairs~\cite{Ema:2020ggo}. Even in this case, this frequency still represents the break of the spectral shape and can become a finger print.
The detection of the break of the spectrum would be a smoking gun proving the existence of  inflation.
Also, it would tells us the energy scale of inflation because 
the break frequency is determined by the Hubble parameter at the end of inflation:
\begin{equation}
f_{*} = \frac{H_{{\rm end}}}{2\pi} \ ,  \label{Hend}
\end{equation}
where $f_{*}$ and $H_{{\rm end}}$ are the break frequency and the Hubble parameter at the end of 
inflation.
After inflation, 
this cutoff frequency is red-shifted due to the expansion of the universe to the frequency $f_{{\rm break}}$:
\begin{equation}
  f_{{\rm break}} = \frac{a_{{\rm end}}}{a_{0}} f_{*} \ ,  \label{fcut}
\end{equation}
where $a_{{\rm end}}$ and $a_{0}$ are the scale factor at the end of inflation and today, respectively.

In general, an inflaton field begins to oscillate around the bottom of the potential after inflation and 
particle production occurs~\cite{ABBOTT198229,DOLGOV1982329}.
In this reheating phase, the evolution of the universe mimics that of matter dominated phase 
approximately~\cite{Kolb:1990vq}.
As a result of (light) particle production during the reheating,
radiation dominated phase follows.
Therefore, the Hubble parameter evolves during the reheating as
\begin{equation}
  \frac{H_{{\rm reh}}}{H_{{\rm end}}} = \left( \frac{a_{{\rm reh}}}{a_{\rm end}} \right)^{-3/2} \  , \label{Hreh}
\end{equation}
where $a_{{\rm reh}}$ is the scale factor at the end of the reheating phase and thus 
at the beginning of the radiation dominated phase.
The Hubble parameter at the end of the reheating phase can be parameterized by the 
reheating temperature, $T_{{\rm reh}}$:
\begin{equation}
  H_{{\rm reh}}^{2} = \frac{\pi^{2} g(T_{{\rm reh}})}{90 M_{\rm pl}^{2}} T_{{\rm reh}}^{4} \ ,  \label{rad}
\end{equation}
where $g (T)$ stands for effective degrees of freedom for the energy density
at a temperature $T$~\cite{Kolb:1990vq,Maggiore:1999vm}.
From Eqs.\,(\ref{Hreh}) and (\ref{rad}), 
one can estimate the expansion rate $a_{{\rm end}}/a_{{\rm reh}}$ by 
using parameters, $H_{{\rm end}}, T_{{\rm reh}}$ and $g(T_{{\rm reh}})$.

In order to estimate the ratio $a_{{\rm reh}} / a_{0}$, we use the fact that 
the entropy $S$ conserves in an adiabatic universe:
\begin{equation}
  S \propto g_{s}(T) a(T)^{3} T^{3}  = {\rm const.} \ .  \label{entropy}
\end{equation}
$g_{s} (T)$ represents effective degrees of freedom for the entropy
at a temperature $T$~\cite{Kolb:1990vq,Maggiore:1999vm}.

Using Eqs.\,(\ref{Hend})-(\ref{rad}), we eventually obtain
\begin{equation}
  f_{{\rm break}} \simeq 1.4 \times 10^{7}\,{\rm Hz}~
       \left( \frac{H_{{\rm end}}}{10^{-5} M_{\rm pl}}  \right)^{1/3}   
       \left( \frac{ T_{{\rm reh}}}{10^{13} {\rm GeV}}  \right)^{1/3}  
       \ .  \label{fcut1}
\end{equation}
Here we used the values for each parameter:
$g_{s}(T_{{\rm reh}}) = g(T_{{\rm reh}})$, 
$g_{s}(T_{0}) \simeq 3.36$,
$T_{0} \simeq 2.73 {\rm K}$.
Suppose that we observe gravitational waves at the break frequency.
Then, because the amplitude also depends on $H_{{\rm end}}$ and $T_{{\rm reh}}$
(see Eq.\,(\ref{eneden})),
it means we can determine not only the energy scale of inflation 
but also the reheating temperature simultaneously.

Finally, we mention that
the maximum value of the break frequency is given by the case of the instantaneous reheating,
i.e., $H_{{\rm end}} = H_{{\rm reh}}$, that is
\begin{equation}
  f_{{\rm break}} \simeq 1.0 \times 10^{8}\,{\rm Hz}~
       \left( \frac{200}{g(T_{{\rm reh}})}  \right)^{1/12}   
       \left( \frac{ H_{{\rm end}}}{10^{-5} M_{\rm pl}}  \right)^{1/2}  
       \ .  \label{fcut2}
\end{equation}
\section{Hubble parameter around the break}\label{HuEvo}
As mentioned in the section \ref{sec2}, the spectrum of primordial gravitational waves is 
determined by the Hubble parameter.
However, gravitational waves around the break frequency ($\ref{fcut2}$) was produced just before 
the end of inflation where the slow roll approximation gets worse.
Therefore, we need a prescription away from the slow roll approximation to discuss the spectrum of primordial gravitational waves around the break frequency.
In this section we give an analytic expression for the Hubble parameter near the end of inflation.

Let us consider an inflationary universe which can be characterized by the scale factor $a(t)$ defined by
\begin{equation}
  ds^{2} = -dt^{2} + a^{2}(t) d \bm{x}^{2} \ .
\end{equation}
Exponential expansion of the universe is driven by an inflaton $\phi(t)$ whose potential 
is $V(\phi)$.
Then, one can obtain Einstein equations:
\begin{eqnarray}
  H^{2} &=& \frac{1}{3 M_{\rm pl}^{2}} \left[ \frac{1}{2} \dot{\phi}^{2} + V \right] \ , \label{H2} \\
  \dot{H} &=& - \frac{1}{2 M_{\rm pl}^2} \dot{\phi}^{2}  \ , \label{Hdot}
\end{eqnarray}
where a dot represents derivative with respect to $t$ 
and $H = \dot{a}/{a}$ is the Hubble parameter.
It is more convenient to express the equations by taking $\phi$ as a time 
variable~\cite{Salopek:1990jq,Lidsey:1991zp}.
Dividing Eq.\,(\ref{Hdot}) by $\dot{\phi}$ yields
\begin{equation}
 H'(\phi) = - \frac{1}{2 M_{\rm pl}^2} \dot{\phi} \ ,   \label{Hd}
\end{equation}
where a dash stands for derivative with respect to $\phi$.
Note that we assume $\dot{\phi} > 0$ ($H' < 0$) throughout our discussion.
Using Eq.\,(\ref{Hd}) to eliminate $\dot{\phi}$ in Eq.\,(\ref{H2}), one obtains
\begin{equation} 
   H'^{2}(\phi)  =  \frac{3}{2 M_{\rm pl}^{2}} H^{2}(\phi)  -  \frac{1}{2M_{\rm pl}^{4}} V(\phi)  \ . 
   \label{Ha}
\end{equation}
This is the equation for the Hubble parameter with the time variable $\phi$.
It should be stressed that the Hubble parameter $H(\phi)$ and the potential $V(\phi)$ are independent.
Indeed, one can not neglect $\dot{\phi}$ in Eq.\,(\ref{H2}) near the end of inflation
and thus $H$ and $V$ are not related directly in contrast to the case of the slow roll 
approximation~\cite{Liddle:1994dx}.
However at the break point, they are related directly as we will see soon.

Differentiating the above equation with respect to $\phi$, we obtain
\begin{equation}
  H''(\phi) =   \frac{1}{2 H'}   
                \left[ \frac{3 H H'}{M_{\rm pl}^{2}}  -  \frac{1}{2M_{\rm pl}^{4}} V'
                \right] \ .  \label{Ha2}
\end{equation}
Here, $H'$ is given by $H$ and $V$ in Eq.\,(\ref{Ha}), so that
$H''$ can be regarded as a function of $H, V$ and $V'$.
%
%
Similarly, differentiating Eq.\,(\ref{Ha2}), we get $H'''(\phi)$ 
as a function of $H, V, V'$ and $V''$.
One can repeat the procedure and $n$ derivatives of the Hubble parameter, $H^{(n)}$, can be written as a function of 
$H, V, V^{1}, \cdots , V^{(n-1)}$.
The formal expression is
\begin{widetext}
\begin{equation}
  H^{(n)} = \frac{1}{2 H^{(1)}} 
            \Bigg[  - \sum\limits_{r=1}^{n-2} \binom{n-1}{r} H^{(r+1)} H^{(n-r)} 
                    + \frac{3}{2 M_{\rm pl}^{2}} 
                    \sum\limits_{r=0}^{n-1} \binom{n-1}{r} H^{(r)} H^{(n-r-1)} 
                   - \frac{1}{2M_{\rm pl}^{4}} V^{(n-1)}  
                   \Bigg] \  ,  \label{Hn}
\end{equation}
\end{widetext}
for $n \geq 3$ and $\binom{n}{r} = \frac{n!}{r!(n-r)!}$.

We now define  the break point $\phi_{0}$ by the equation 
\begin{eqnarray}
\epsilon (\phi_0) \equiv 2M_{\rm pl}^{2}\frac{H^{\prime 2}}{H^2}=1 \ .
\end{eqnarray}
On the break point, the Hubble parameter and the potential are linked, 
\begin{equation}
  H(\phi_{0}) = \sqrt{\frac{V(\phi_{0})}{2 M_{\rm pl}^{2}}}  \ .  \label{brerela}
\end{equation}
Then, we can expand the Hubble parameter:
\begin{eqnarray}
  H(\phi ) 
&=& \sum_{n=0}^{\infty} \frac{1}{n!} H^{(n)}(\phi_{0}) \left( \phi - \phi_{0} \right)^{n} \nonumber \\
           &=& \sum_{n=0}^{\infty} 
               \frac{1}{n!} H^{(n)} \left(V(\phi_{0}), \cdots, V^{(n-1)}(\phi_{0}) \right) \nonumber \\
           &\ & \quad \quad \times \left( \phi - \phi_{0} \right)^{n}  \ .  \label{Hexp} 
\end{eqnarray}
We used Eqs.\,(\ref{Hn}) and (\ref{brerela}) to get the second equality.
Eq.\,(\ref{Hexp}) shows that the evolution of the Hubble parameter, which is needed to calculate 
the spectrum of primordial gravitational waves, can be described 
by the inflaton potential and its derivatives, which are evaluated at $\phi_{0}$.
\begin{figure}[h]
\centering
  \includegraphics[width=8cm]{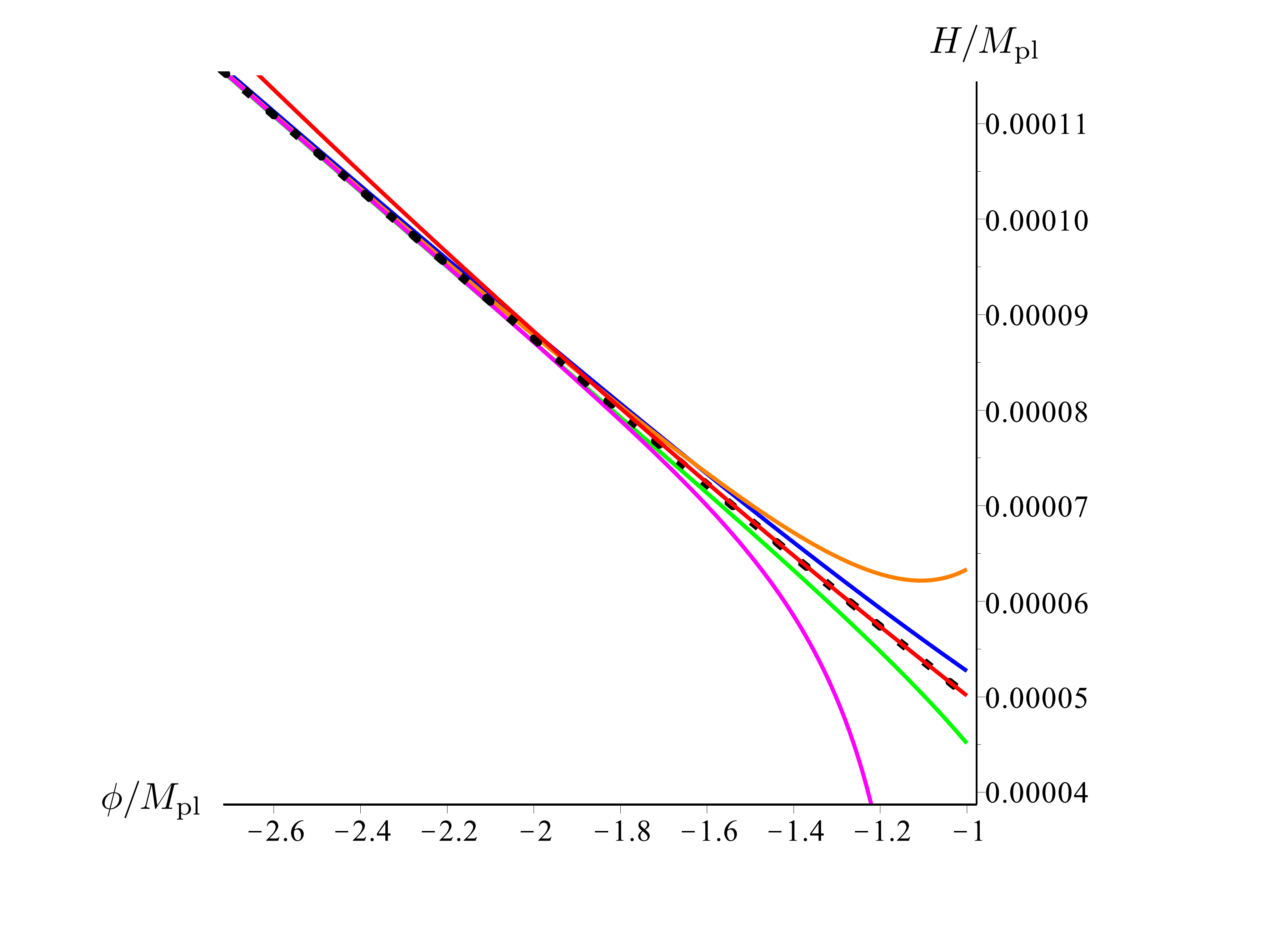}
  \caption{The Hubble parameters near the end of inflation in the case of $V(\phi)\propto \phi^{2}$ are depicted.
  $\phi / M_{\rm pl} = -1$ corresponds to the end of inflation where a slow roll parameter becomes unity. 
  The red line represents Eq.\,(\ref{Hexp}) with $n=2$ and expanded from the end of inflation, i.e.,
  $\phi_{0} / M_{\rm pl} = -1$.
  The black dotted line is a numerical solution of Eq.\,(\ref{Ha}).
  The blue, green, orange, pink lines shows approximated solutions of 1st, 2nd, 3rd and 4th slow roll 
  approximations, respectively~\cite{Liddle:1994dx}. 
  }
  \label{Hubble}
\end{figure}
In Fig.\,(\ref{Hubble}), we plot Eq.\,(\ref{Hexp}) (red line) at second order, $n=2$, 
from the end of inflation with the exact numerical value (black doted line) and results from the slow roll approximation~\cite{Liddle:1994dx} (blue, green, orange, pink lines).
Note that although we chose the potential form $V(\phi) \propto \phi^{2}$,
our discussion can be applied to any form of the potential in parallel.
We see that our formula (\ref{Hexp}) to the second order expansion ($n=2$) 
shows good agreement with the numerical one 
between, $\phi = -2 M_{\rm pl} \sim - M_{\rm pl}$, where the slow roll approximation gets worse.
It should be mentioned that the slow roll approximation seems to break down near the end of inflation, 
namely higher order corrections do not make the result better.
Therefore, Eq.\,(\ref{Hexp}) is complementary to the slow roll approximation%
\footnote{
In~\cite{Liddle:1994dx}, the Pad\'{e} approximant was used and a better result was obtained compared with
simple slow roll approximations.
However, even in that case, the deviation is unavoidable around the end of inflation.
}.
\section{Primordial gravitational waves around the break}\label{PSpe}
In the previous section, we gave an analytic expression (\ref{Hexp}) to describe the evolution of 
the Hubble parameter near the end of inflation.
It can be used to derive the spectrum of primordial gravitational waves, which is expected 
to be given by Eq.\,(\ref{spe}).
In this section,
we first discuss the power spectrum of primordial gravitational waves 
at the spectral break.
Next, we
explicitly calculate the spectral index of the spectrum at the end of inflation.
Finally, we give a spectrum around the spectral break at present.
\subsection{Spectrum at the end of inflation}
During inflation, the equation of motion for the canonically normalized tensor perturbation
$v_{k,\sigma}$ ($\sigma$ stands for degrees of freedom of the polarization)
), which is related to the tensor perturbation $h_{k,\sigma}$,
 $v_{k,\sigma} = \frac{M_{{\rm pl}} a}{ 2}h_{k,\sigma}$, is
given by
\begin{equation}
    \frac{d^{2} v_{k,\sigma}(\tau)}{d\tau^{2}}
    +\left( k^{2} - \frac{\frac{d^{2} a}{d \tau^{2}}}{a} \right) 
    v_{k,\sigma}(\tau) 
                   = 0 \ ,  \label{canoni}
\end{equation}
where $\tau = \int da 1/aH$ is the conformal time.
%
%
Then, one can approximate the term in the parenthesis,
\begin{equation}
  \frac{\frac{d^{2} a}{d \tau^{2}}}{a} 
    = 2 a^{2} H^{2} \left( 1 - \frac{1}{2}\epsilon \right)  
    \simeq 2 a^{2} H^{2} \ .  \label{yuragi}
\end{equation}
Here we used the fact that the first slow roll parameter $\epsilon$ is 
always less than unity even around the end of inflation.
Therefore, we do not require that higher order slow roll parameters
are less than unity, which is violated as we saw in Sec.III.
Then, moreover, the conformal time can be rewritten with integration by parts
and approximated as
\begin{equation}
  \tau \simeq -\frac{1}{aH}\frac{1}{1-\epsilon} \simeq -\frac{1}{aH} \ .
\end{equation}
Therefore, Eq.\,(\ref{canoni}) can be approximated:
\begin{equation}
    \frac{d^{2} v_{k,\sigma}(\tau)}{d\tau^{2}}
    +\left( k^{2} -  \frac{2}{\tau^{2}} \right) 
    v_{k,\sigma}(\tau) 
                   = 0 \ .  \label{usual}
\end{equation}
We notice that this is a usual equation for tensor perturbations in the slow roll approximation.
The point is that the first slow roll parameter $\epsilon$ is always less than 
unity although more higher order slow roll parameters could be larger than 
unity around the inflation end as indicated in Sec.\,\ref{HuEvo}.
From Eq.\,(\ref{usual}),
we can get the dimension less power spectrum for $h_{k,\sigma}$
at the end of inflation, i.e. Eq.\,(\ref{spe}).
Replacing $H$ in the power spectrum by the analytic expression (\ref{Hexp}), 
we can calculate it analytically.
The obtained power spectrum is more accurate than the totally slow roll approximated one 
as we will see in Sec.\,\ref{spepre}.
\subsection{Spectral index}\label{spein}
In this subsection, we analyze the spectral index of the tensor perturbation,
which would be an important observable, with our analytic formulation (\ref{Hexp}).
We define the spectral index $n_{T}$ at the end of inflation by
\begin{equation}
  n_{T}   = \frac{d\ln(P_{h}(k))}{d\ln k} 
          = \frac{d\ln(P_{h}(k))}{d\ln a} \times \frac{d\ln a}{d\ln k}  \ .  \label{ind}
\end{equation}
Together with Eq.\,(\ref{spe}), we have
\begin{equation}
  \frac{d\ln(P_{h}(k))}{d\ln a} = 2 \frac{d\ln H}{d\ln a} = - \frac{4 M_{\rm pl}^{2} H'^{2}}{H^{2}} \ .
\end{equation}
On the other hand, 
\begin{eqnarray}
  \frac{d\ln a}{d\ln k} &=& \left( \frac{d\ln k}{d\ln a} \right)^{-1}   
             = \left( 1 +  \frac{d\ln H}{d\ln a}  \right)^{-1}     \nonumber \\
             &=& \left( 1 - \frac{2 M_{\rm pl}^{2} H'^{2}}{H^{2}}  \right)^{-1}  \ .   \label{dadk}
\end{eqnarray}
Note that we used the relation $k=aH$.
Therefore, from Eqs.\,(\ref{ind})-(\ref{dadk}), we obtain
\begin{equation}
  n_{T} = - \frac{ 4 M_{\rm pl}^{2} H'^{2}(H,V)}{ H^{2} - 2 M_{\rm pl}^{2} H'^{2}(H,V)}  \ ,  \label{index}
\end{equation}
where we explicitly indicated that $H'$ is a function of $H$ and $V$ through Eq.\,(\ref{Ha}).

Observing that the scale factor can be expressed by an integral,
\begin{equation}
 \ln a = \int  - \frac{H}{2 M_{\rm pl}^2 H'(H,V)}  d\phi \ ,
\end{equation}
one can rewrite the wave number $k$ (or the frequency $f$) as
\begin{equation}
 k =  \exp \left(\int  - \frac{H}{2 M_{\rm pl}^2 H'(H,V)}  d\phi \right)  H \quad  (= 2 \pi f  )   \ .   \label{k}
\end{equation}
We see that Eqs.\,(\ref{index}) and (\ref{k}) are expressed by $H$ and $V$.
Replacing $H$ in the equations by the analytic expression (\ref{Hexp}), 
we can calculate the spectral index against $k$ semi-analytically. 
%
%
%
%
Note that the spectral index diverges at the end of inflation because the denominator 
of Eq.\,(\ref{index}) is zero at there. 
This clearly shows the existence of the break point.

An advantage of using the analytic expression is that 
one can extract the important information from the formula.
For example, if one expands $n_{T}$ in powers of $(\phi-\phi_{0})$, 
we have
\begin{widetext}
\begin{eqnarray}
  n_{T} =  \left( \frac{\sqrt{2}}{M_{\rm pl}} - \frac{V'(\phi_{0})}{V(\phi_{0})}  \right)^{-1} (\phi - \phi_{0})^{-1} +
          \frac{   \frac{4}{M_{\rm pl}^{2}} - \frac{9}{\sqrt{2}M_{\rm pl}} \frac{V'(\phi_{0})}{V(\phi_{0})}
                + 2 \left( \frac{V'(\phi_{0})}{V(\phi_{0})} \right)^{2}
                + \frac{1}{2} \frac{V''(\phi_{0})}{V(\phi_{0})}}
               { \left( \frac{\sqrt{2}}{M_{\rm pl}} - \frac{V'(\phi_{0})}{V(\phi_{0})}  \right)^{2} } 
                 + \cdots .
\end{eqnarray}
\end{widetext} 
We note that the denominators, 
$(\sqrt{2}/M_{\rm pl} - V'(\phi_{0})/V(\phi_{0})$, are not zero and, furthermore, not
necessarily small because the slow roll approximations is broken at
$\phi_{0}$.  This analytic expression can be used as a fitting function
for data analysis.  Since the fitting function includes information of
the inflaton potential at the break point, it is useful to discriminate
inflation models.  Indeed, for example, the sign of $V''(\phi_{0})$ is
different from a model to a model, so that determining the sign of
$V''(\phi_{0})$ from observations is important to distinguish inflation
models. 
It should be mentioned that although
the term proportional to $V''$ may be subdominant in the spectral index,
this does not mean that we can not observe it.
Indeed, if we have high enough sensitivity to the spectral index, we can still extract the 
contribution from the term proportional to $V''$.
Therefore, our analytic expression could be useful.
\subsection{Spectrum at present} \label{spepre}
In this subsection, we give an example of the primordial gravitational wave spectrum 
around the spectral break at present.
The dimensionless power spectrum around the end of inflation $P_{h}(k)$
is given by Eq.\,(\ref{spe}).
After inflation, in the super horizon regime, $k < aH$, the power spectrum keeps the initial amplitude.
When a mode $k$ reenter the Hubble horizon, namely in the sub horizon regime $k > aH$,
the power spectrum starts to decay proportional to $a^{-2}$.
It means the spectrum at present depends on the background evolution of the universe through the scale factor.

It would be useful to define the energy density parameter instead of the power spectrum 
and evolve it to the present time:
\begin{equation}
\frac{d \ln \Omega_{{\rm GW}}^0}{d \ln k} = \frac{k^{2}
P_{h}(k)}{12 a_0^2 H_{0}^{2}} \times \left(\frac{a_{{\rm reenter}}(k)}{a_{0}} \right)^2 \ . 
\end{equation}
where $H_{0} = 70 \,{\rm km}/{\rm s\,  Mpc}$ is the Hubble parameter at present and 
$a_{{\rm reenter}}$ is the scale factor when a mode reenters the Hubble horizon, 
i.e., $k=a_{{\rm reenter}} H$.

We now assume that, as was done in Sec.\,\ref{sec2}, 
a reheating phase where the inflaton oscillates around the bottom of its potential
follows after inflation.
In the phase, the Hubble parameter evolves as $H \propto a^{-3/2}$ and then 
$a_{{\rm reenter}}(k) \propto k^{-2}$.
Therefore, the energy density parameter would proportional to $k^{-2}$.
More explicitly, the energy density parameter for the modes which
reenter during the \rm{inflaton oscillation} is%
\footnote{
We have not considered the effect of anisotropic stress of neutrinos~\cite{Watanabe:2006qe,Kuroyanagi:2008ye} here.
It would give an overall suppression factor to Eq.\,(\ref{eneden}).
}
%
\begin{eqnarray}
  \frac{d \ln \Omega_{{\rm GW}}^0}{d \ln f}  &=& 
    \frac{\pi^{2} f_{{\rm break}}^{2}  P_{h}(f)}{3 H_{0}^{2}}
    \left( \frac{f}{f_{{\rm break}}}  \right)^{-2} \nonumber \\
    &&   \times 
           \left( \frac{g_{s}(T_{0}) T_{0}^{3}}{g_{s}(T_{{\rm reh}}) T_{{\rm reh}}^{3}} \right)^{2/3}
           \left( \frac{H_{{\rm reh}}}{H_{{\rm end}}} \right)^{4/3} \ .    \label{eneden}
\end{eqnarray}
%
Note that $H_{{\rm reh}}$ is related to $T_{{\rm reh}}$ by Eq.\,(\ref{rad})
and $f_{{\rm break}}$ is determined by $T_{{\rm reh}}$ and $H_{{\rm end}}$ in Eq.\,(\ref{fcut1}).
Therefore, for a set of free parameters $T_{{\rm reh}}, H_{{\rm end}}$, one can calculate 
Eq.\,(\ref{eneden}) immediately with the semi-analytic formula
we investigated in the previous sections.
\begin{figure}[t]
\centering
  \includegraphics[width=9.2cm]{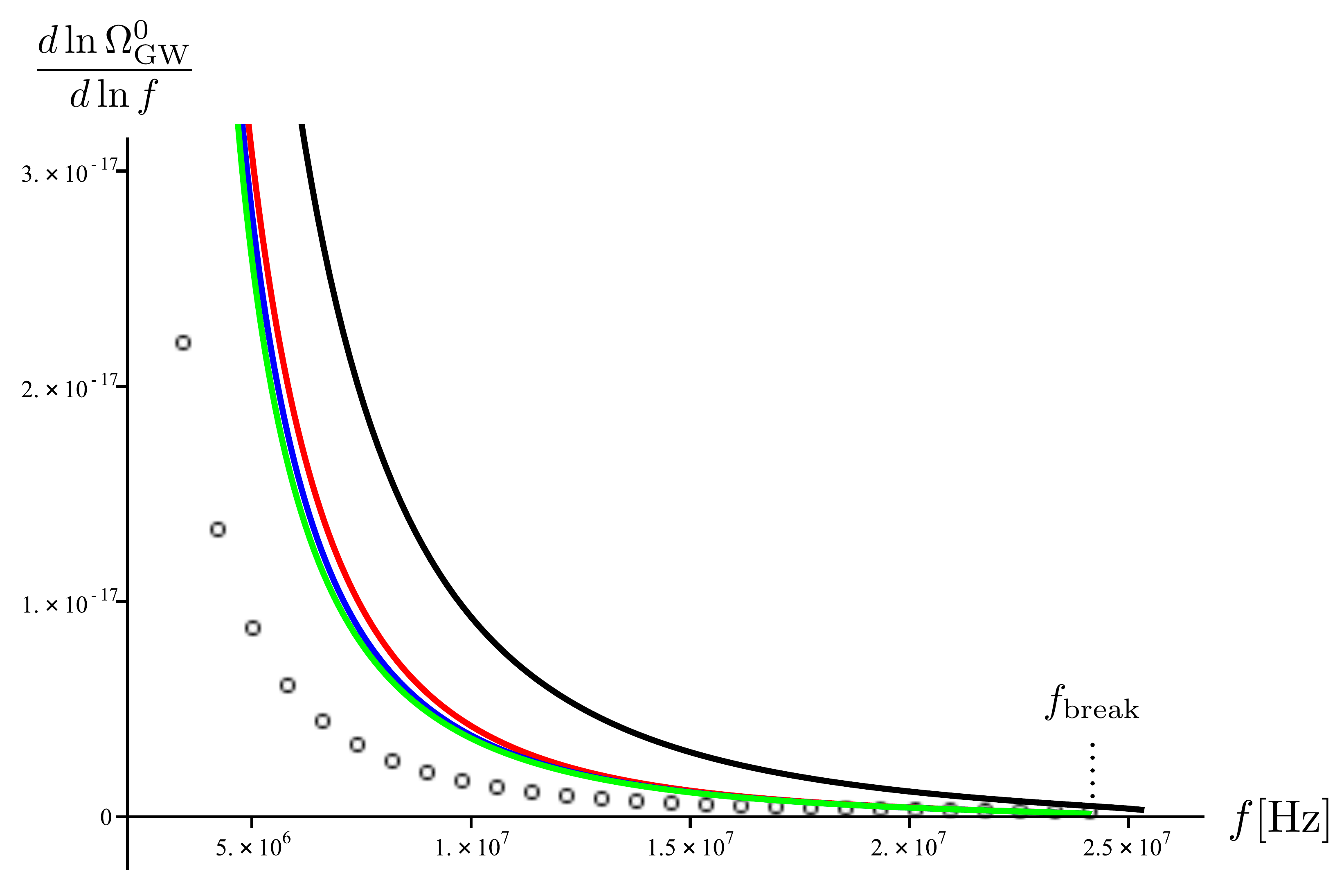}
  \caption{The energy density parameter (\ref{eneden}) 
     around $f_{{\rm break}}$ given by Eq.\,(\ref{fcut1}) is depicted.
     The black dot line is for a numerical solution where we solved Eq.\,(\ref{yuragi}) numerically.
     The red, blue and green lines represent our semi-analytic method with $n=2$, $n=4$ and $n=6$ 
     in Eq.\,(\ref{Hexp}), respectively.
     The black line is the result of the 1st order slow roll approximation.
     }  \label{present}
\end{figure}
Let us give an example of the spectrum in the case of $V(\phi) = m^{2} \phi^{2}$ where 
we set $T_{{\rm reh}} = 10^{13}$ GeV and $m = 10^{-4} M_{{\rm pl}}$
(the corresponding value of the Hubble at the end is, $H_{{\rm end}} = 5.0 \times 10^{-5} M_{\rm pl}$). 
The result is depicted in Fig.\,\ref{present}.
We see that our semi-analytic method (red, blue and green lines) well fit the numerical result (black dot line) 
around the spectral break where the slow roll approximation (black line) gets worse.
Note that we used the 1st order slow roll approximation for the black line because
higher order collections make the result worse around the spectral break (see Fig.\,\ref{Hubble}).
Fig.\,\ref{present} shows that our formulation gives more accurate result than the 
slow roll approximation around the spectral break.
Therefore, our formulation with the slow roll approximation (expansion) would enable us to complete 
the spectrum in full frequency range in analytic ways.
\section{Conclusion}
We studied a spectral break of primordial gravitational
waves as a fingerprint of the end of inflation.
Indeed, the amplitude and the
frequency on the break can tell us the energy scale of inflation
and the reheating temperature simultaneously. 
This spectral break scale may exist around $10^7$ Hz
(see Eq.\,(\ref{fcut1})) and thus the theoretical investigation is crucial for new gravitational wave detectors developed 
intensively~\cite{Domcke:2020yzq,Ito:2020wxi,Ito:2019wcb,Li:2009zzy,Li:2008qr}.

We gave an analytic formula for describing the evolution of the Hubble parameter near the end of inflation in Sec.\,\ref{HuEvo}.
Because the slow roll approximation breaks down near the end of inflation,
our analytic formulation is useful and complementary to it.
In Sec.\,\ref{PSpe}, 
we first gave an analytic expression of the power spectrum
of tensor perturbations at the end of inflation.
We also semi-analytically
investigated the spectral index of the spectrum at 
the end of inflation in Sec.\,\ref{spein}. 
It was discussed how to extract important information of the inflaton potential 
from the observation of the spectral index.
In Sec.\,\ref{spepre}, we presented an example of the spectrum around the break point at present in Fig.\,\ref{present}
for specific values of parameters in Eq.\,(\ref{eneden}).

It should be mentioned that probing the reheating temperature with 
primordial gravitational waves was also proposed in~\cite{Nakayama:2008ip,Nakayama:2008wy,Kuroyanagi:2014qza}.
However, the paper focused on a bend of the spectrum due to the transition of background spacetime
from the reheating to the radiation dominant phase.
Then, the possibility to observe the bend with gravitational wave interferometers was discussed.
This situation is realized when the reheating temperature is low enough.
In the present paper, we intend to observe the spectral break due to the end of inflation and then
higher reheating temperature is favorable compared with the above case.
\begin{acknowledgments}
A.\,I. and J.\,S. was supported by JSPS KAKENHI Grant Numbers
JP17H02894, JP17K18778.  M.\,Y. is supported in part by JSPS
Grant-in-Aid for Scientific Research Numbers 18K18764, Mitsubishi
Foundation, and JSPS Bilateral Open Partnership Joint Research Projects.
\end{acknowledgments}
\bibliography{cutoff}

\end{document}